\documentclass[letterpaper, 10 pt, conference]{ieeeconf}  

\IEEEoverridecommandlockouts                              

\usepackage{algorithm}
\usepackage[noend]{algpseudocode}
\usepackage{amsmath}
\usepackage{amssymb}
\usepackage{array}
\usepackage[backend=bibtex,sorting=none]{biblatex}
\usepackage{booktabs}
\usepackage{caption}
\usepackage{dblfloatfix}
\usepackage{enumerate}
\usepackage{gensymb}
\usepackage{graphicx}
\usepackage{multirow}
\usepackage{indentfirst}
\usepackage{tabularx}
\usepackage{threeparttable}
\usepackage[utf8]{inputenc}
\usepackage{subcaption}
\usepackage{mathtools}
\DeclarePairedDelimiter{\floor}{\lfloor}{\rfloor}
\usepackage{color}

\title{\LARGE \bf
Fast Trajectory Planning for Multiple Quadrotors \\using Relative Safe Flight Corridor
}

\author{Jungwon Park$^{1}$ and H. Jin Kim$^{1}$
\thanks{$^{1}$Jungwon Park is with the Department of Mechanical and Aerospace Engineering, Seoul National University, Seoul,         South Korea
        {\tt\small qwerty35@snu.ac.kr}}%
\thanks{$^{1}$H. Jin Kim is with the Department of Mechanical and Aerospace Engineering, Seoul National University, Seoul,         South Korea
       {\tt\small hjinkim@snu.ac.kr}}%
}

\begin{document}

\maketitle
\thispagestyle{empty}
\pagestyle{empty}

\begin{abstract}
This paper presents a new trajectory planning method for multiple quadrotors in obstacle-dense environments. 
We suggest a relative safe flight corridor (RSFC) to model safe region between a pair of agents, and it is used to generate linear constraints for inter-collision avoidance by utilizing the convex hull property of relative Bernstein polynomial. 
Our approach employs a graph-based multi-agent pathfinding algorithm to generate an initial trajectory, which is used to construct a safe flight corridor (SFC) and RSFC. We express the trajectory as a piecewise Bernstein polynomial and formulate the trajectory planning problem into one quadratic programming problem using linear constraints from SFC and RSFC. The proposed method can compute collision-free trajectory for 16 agents within a second and for 64 agents less than a minute, and it is validated both through simulation and indoor flight test.

\end{abstract}

\section{INTRODUCTION}
Multi-agent systems consisting of micro aerial vehicles (MAVs) are  receiving attention from many industrial domains due to their agility, mobility, and applicability. To maximize their capabilities for various missions such as cooperative surveillance \cite{saska2016swarm} and transportation \cite{kim2017motion}, it requires to generate safe trajectories for multiple quadrotors in a complex environment within a short time.
However, it has been challenging to efficiently formulate constraints to avoid obstacles and other agents. Furthermore, deadlock may happen if agents are packed in a narrow space. In this paper,
we focus on an efficient planning method in terms of both cost and computation time which generates safe, dynamically feasible trajectories in an obstacle-dense environment without deadlock.

One  popular approach to generate multi-agent trajectories is a centralized optimization method
In \cite{mellinger2012mixed}, constraints for collision avoidance are reformulated in integer constraints for mixed-integer quadratic programming (MIQP). However, it requires over 500--1000 seconds to optimize the trajectory of 2--4 agents due to the computational complexity of the MIQP.
In \cite{augugliaro2012generation}, sequential convex programming (SCP) is proposed to replace non-convex constraints with convex ones. SCP shows good performance when planning a small number of quadrotors, but it is intractable for a large team and complex environment.
The authors of \cite{robinson2018efficient} suggest nonlinear programming (NLP) which combines sequential planning  to deal with nonlinear constraint directly. This use of sequential planning method allows to achieve better scalability, but it has a limitation that no feasible solution can be found for a crowded situation.

A decentralized method also has been 
considered to reduce total planning time by distributing the computational load. 
Approaches based on LQR-obstacle \cite{bareiss2013reciprocal} and buffered Voronoi cells \cite{zhou2017fast} show that they can generate a collision-free path in real time. However, such distributed methods are not able to guarantee the completeness, no deadlock. In \cite{dai2017distributed} and \cite{wang2014synthesis}, distributed model predictive control (DMPC) is proposed to optimize the control input instead of a piecewise polynomial path, but they do not consider obstacles.

In single quadrotor path planning, many researchers have adopted a safe flight corridor (SFC) to model free space in a map. SFC is composed of connected convex sets, and it can be represented as linear inequality constraints for obstacle avoidance in quadratic programming (QP)  \cite{tang2016safe, liu2017planning, gao2018online}.
In \cite{honig2018trajectory}, SFC is used to separate a safe region of each agent in quadrotor swarm. By resizing SFC iteratively, trajectories of quadrotor swarm can be refined separately without inter-collision. This decoupled iterative optimization method shows good scalability in a maze-like environment, but it requires many iteration steps for the convergence of the overall cost.

\begin{figure}[t]
\centering
\includegraphics[width = 0.7\linewidth]{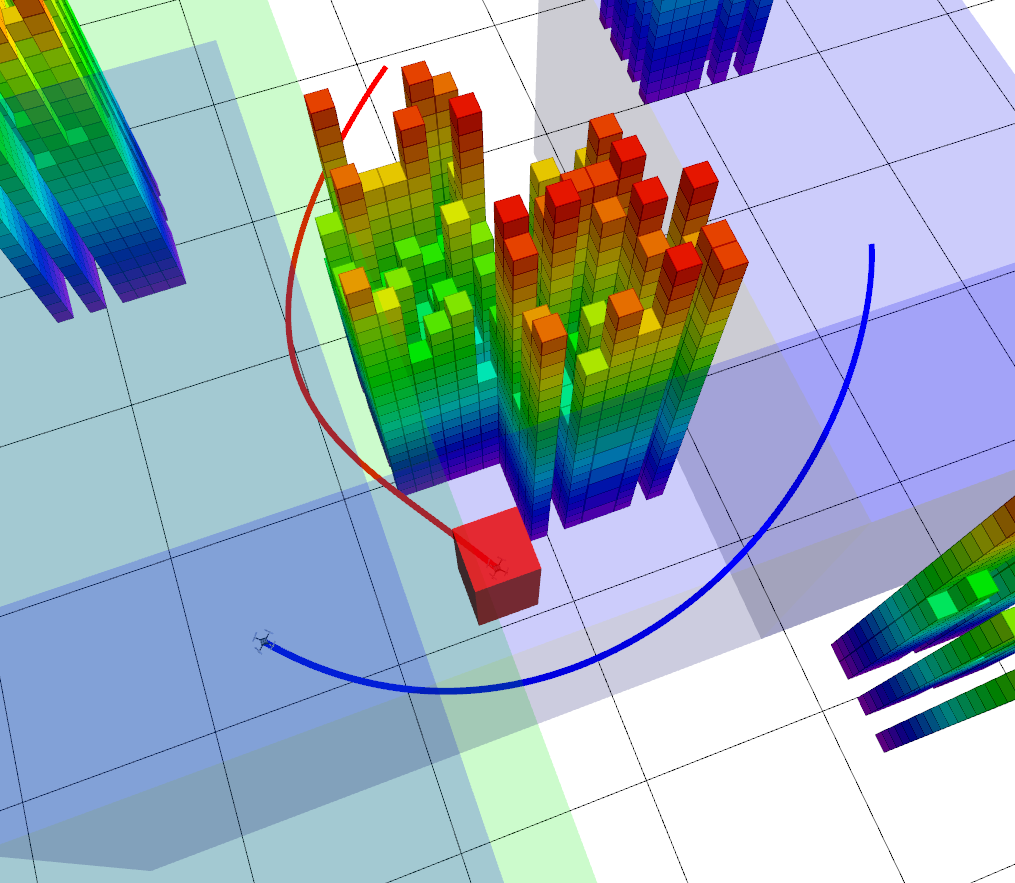}
\caption{
SFC (semi-transparent blue box) and RSFC (semi-transparent green box) used for planning trajectories for two agents (red and blue lines). The trajectory of agent 2 (blue line) is generated within the intersection between SFC and RSFC to avoid obstacles and inter-collision model between two agents (red box)
}
\label{fig:thumnail}
\end{figure}

In this paper, we propose a new centralized multi-agent path planning method that uses a relative safe flight corridor (RSFC) to find a feasible trajectory without any sequential or iterative process.
Similar to SFC, RSFC utilizes a property of Bernstein polynomial to convert non-convex inter-collision avoidance constraints into linear ones as illustrated in Fig. \ref{fig:thumnail}, and it does not need an additional resizing process to find the feasible trajectory.
Thus, our proposed method can optimize a piecewise polynomial trajectory by using QP only once, and it guarantees a feasible solution of QP does not cause a collision and deadlock. 
Recently, distributed planning is receiving much attention due to scalability, yet centralized methods can still provide the quality solution by the efficient formulation using RSFC.

Our main contributions can be summarized as follows.
\begin{itemize}
\item A collision avoidance constraint formulation method using relative safe flight corridor is proposed, which does not require sequential or iterative process.
\item A fast trajectory optimization framework is presented in an obstacle-dense environment, which guarantees  collision- and deadlock-free.
\end{itemize}


This paper is structured as follows.
The problem statement is presented in section \ref{sec: problem statement}. 
In section \ref{sec: method}, we describe the method of multi-agent trajectory planning using relative safe flight corridor.
Experimental results are presented in section \ref{sec: experiments}.
Finally, section \ref{sec: conclusions} contains conclusions.

\section{PROBLEM STATEMENT}
\label{sec: problem statement}
Consider a multi-agent robot system that consists of $N_q$ quadrotors. Each quadrotor is assumed to have different size with radius $r^{1},...,r^{N_q}$ but has the same dynamic limit. We assume that prior knowledge of the free space $\mathcal{F}$ and obstacle $\mathcal{O}$ is given in 3D occupancy map and start, goal point of the $i^{th}$ quadrotor is assigned as $s^i$, $g^i$.

It has been shown that quadrotor dynamics is differentially flat and trajectory can be represented in piecewise polynomials with flat outputs in time $t$ \cite{mellinger2011minimum}. Thus,trajectory of $i^{th}$ quadrotor, $p^i(t)$, can be represented in $M$-segment piecewise polynomials.

In this paper, we aim to generate continuous, smooth trajectory $p^i(t)$ for all $i = 1, ... ,N_q$ which minimizes the integral of the square of the $n^{th}$ derivative and does not collide with any obstacle and other agents. 

\begin{figure*}[t]
\centering
\includegraphics[width = 1\textwidth]{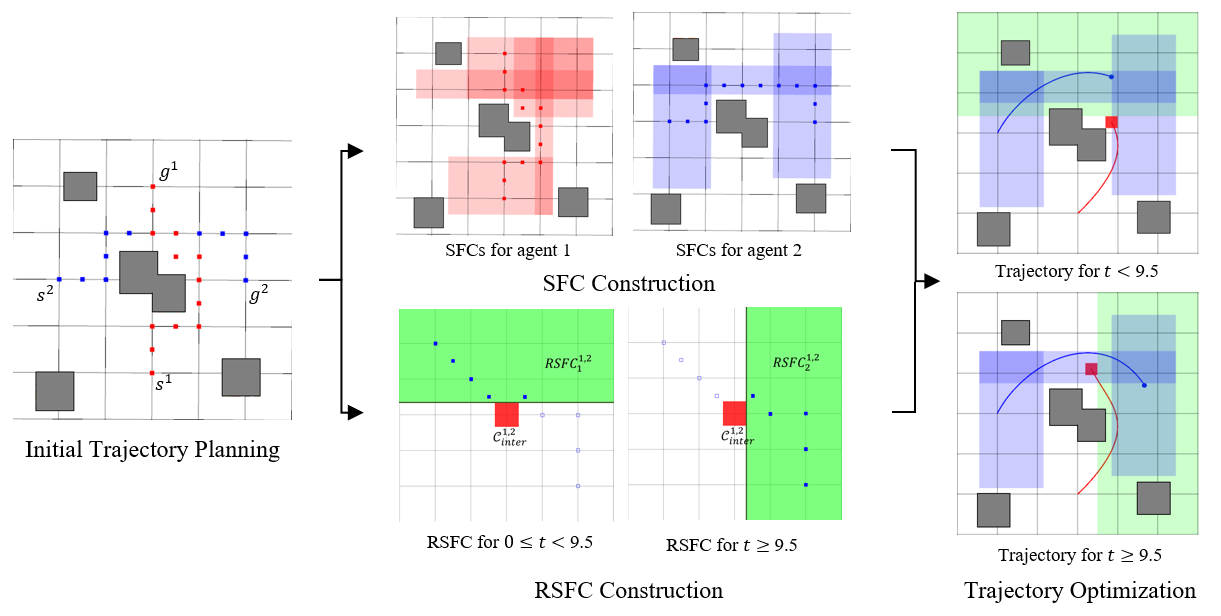}
\caption{Overview of the proposed method. The proposed method can plan the trajectory in the 3D space, but in this example, we plan the trajectory in 2D space for the convenience of explanation. We assigned the start positions for agent 1 (red) and agent 2 (blue) at (0, -2), (0, -2), and goal positions at (0, 2), (2, 0). We convert 3D occupancy map into the 3D grid map and compute discrete initial trajectory by using MAPF algorithm. SFC of each agent (red and blue boxes in SFC construction) is constructed along waypoints of their initial trajectories to prevent obstacle collision. RSFC (green boxes in RSFC construction) is used to confine a relative trajectory between two agents. Finally, SFC and RSFC are translated into linear inequality constraints and QP solver generates the continuous smooth trajectory that satisfies SFC and RSFC constraints.}
\label{fig:structure}
\end{figure*}

\section{METHOD}
\label{sec: method}
The overall structure of our proposed method is depicted in Fig. \ref{fig:structure}.
Our method consists of three steps. First, the discrete path planner plans the initial trajectory using the multi-agent pathfinding (MAPF) algorithm. Then safe flight corridor (SFC) and relative safe flight corridor (RSFC) are constructed based on the initial trajectory. Finally, SFC and RSFC are converted into inequality constraints of quadratic programming (QP) and we obtain the desired trajectory by utilizing the convex hull property of Bernstein polynomial. The detail of each part is described in the following subsections.

\subsection{Initial Trajectory Planning}
When planning the trajectory of a single quadrotor, many researchers have divided the planning process into initial trajectory planning and trajectory refinement, and such two-step method is now being adopted in the multi-agent case \cite{xu2018concurrent, honig2018trajectory}. Inspired by that, we first plan the discrete initial trajectory by using a graph-based MAPF algorithm.

The initial trajectory of the $i^{th}$ quadrotor, $p^i_{init}$, is defined as an array of waypoints that connect start and goal position in a graph. In the MAPF, the cost function is defined as the sum of each trajectory's length.

There have been many researches about MAPF algorithm such as HCA* \cite{silver2005cooperative}, M* \cite{wagner2011m}, conflict-based search (CBS) \cite{sharon2015conflict}. Among them, we choose enhanced CBS (ECBS) \cite{barer2014suboptimal} as a discrete initial trajectory planner because it can find a suboptimal solution in a short time and we can specify the bound of the solution cost. In other words, it is guaranteed that the cost of the trajectory is lower than $c_w \cdot$ optimal cost, where $c_w$ is a user-specified bounding factor.
 
To utilize the graph-based ECBS in our problem, discrete planner translates 3D occupancy map into a 3D grid map. After translation, ECBS computes a discrete initial trajectory that connects start and goal points. If start and goal points are not located on the 3D grid map, then we use the nearest grid points to obtain trajectory and append the start/goal points to both ends. Finally, to calculate relative trajectory between two agents, we match the initial trajectory of all agents as the same length $l_{max}$ by appending each goal point at the end of the trajectory, where $l_{max}$ is the length of the longest initial trajectory. 


\subsection{Safe Flight Corridor Construction}
\label{subsec: SFC}
SFCs of the $i^{th}$ quadrotor, $SFC^{i}_{1},...,SFC^{i}_{M_{s}}$, are defined as convex sets that do not collide with obstacle and are sequentially connected: \\
\begin{subequations}
for $m = 1,...,M_{s}$
\begin{align}
        SFC^i_{m} \oplus \mathcal{C}^i_{obs} \in \mathcal{F}
\label{eq: SFC definition1}
\end{align}
and for $m = 1,...,M_{s}-1$
\begin{align}
        SFC^i_{m} \cap SFC^i_{m+1} \neq \emptyset
\label{eq: SFC definition2}
\end{align}
\end{subequations}
where $\oplus$ is the Minkovski sum and $\mathcal{C}^i_{obs}$ is the obstacle-collision model for the $i^{th}$ quadrotor, which is defined as a sphere with radius $r^i$ representing safety clearance between an obstacle and a quadrotor. 

The trajectory of $i^{th}$ quadrotor is collision-free from obstacle if 
for arbitrary $t \in [0,T]$, there exists  $m \in \{1,...,M_{s}\}$ such that  $p^i(t) \in SFC^i_{m}$, where $T$ is the total flight time.

We construct SFC by the axis-search method. We initialize corridors with a predefined size at each waypoint of the initial trajectory. Then, except the corridors from the start and goal points, we expand them in the previous waypoint direction to connect two sequential convex sets. All waypoints except start and goal points are aligned on the 3D grid map, thus we can achieve the condition (\ref{eq: SFC definition2}) by this method. After that, we expand each corridor in all the other axis-aligned directions until it has a maximum possible free space. Finally, we delete duplicated corridors.

\begin{figure}
    \vspace{1.5mm}
    \centering
    \begin{subfigure}[t]{0.2\textwidth}
        \includegraphics[width=\textwidth]{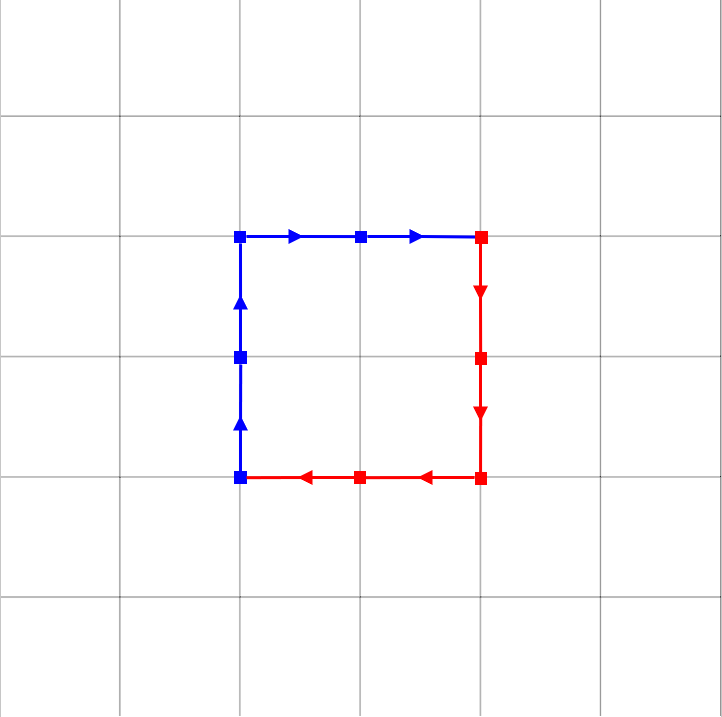}
        \caption{Initial trajectories of the $i^{th}$ (red) and $j^{th}$ (blue) quadrotors.}
        \label{fig:rsfc initial trajectory}
    \end{subfigure}
    ~ 
    \begin{subfigure}[t]{0.2\textwidth}
        \includegraphics[width=\textwidth]{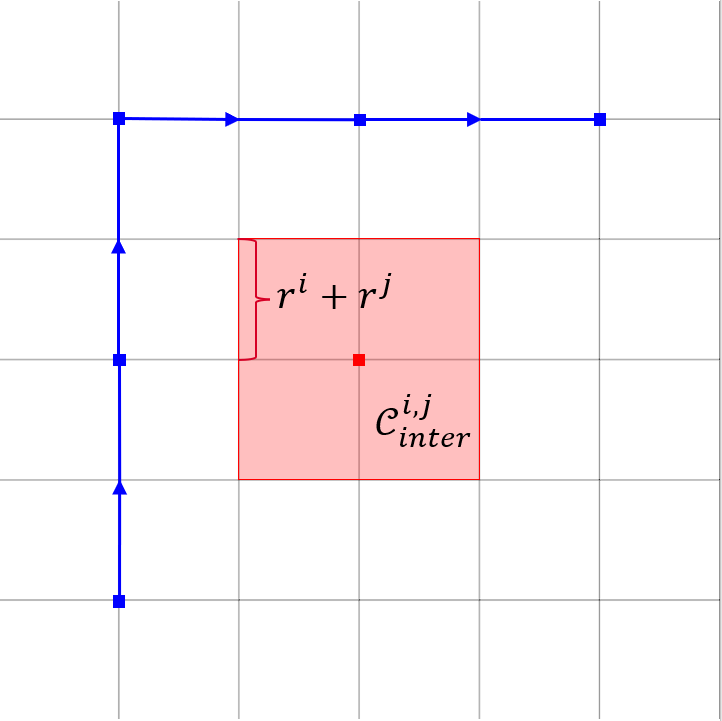}
        \caption{Relative initial trajectory of the $j^{th}$ quadrotor with respect to the $i^{th}$.}
        \label{fig:rsfc relative trajectory}
    \end{subfigure}
    ~ 
    \begin{subfigure}[t]{0.2\textwidth}
        \includegraphics[width=\textwidth]{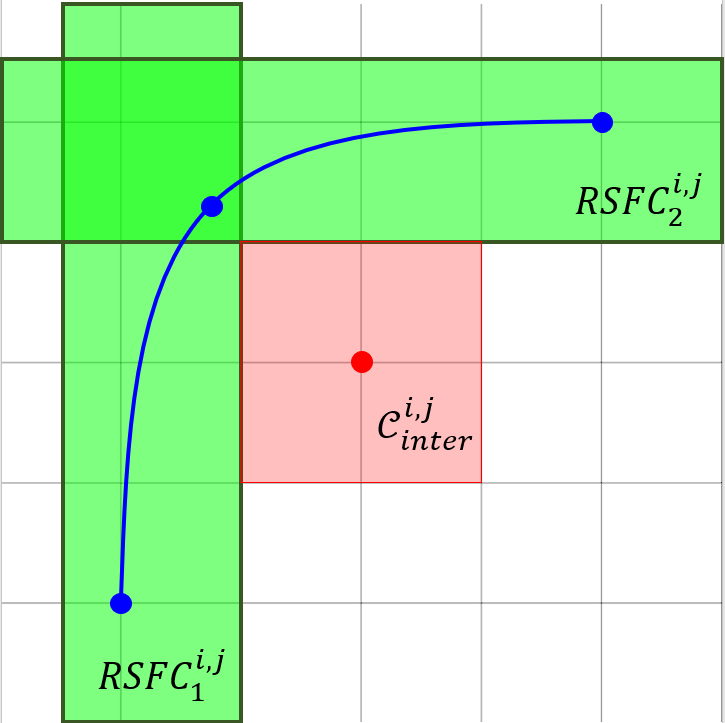}
        \caption{RSFC generated.}
        \label{fig:rsfc rsfc}
    \end{subfigure}
    ~
    \begin{subfigure}[t]{0.2\textwidth}
        \includegraphics[width=\textwidth]{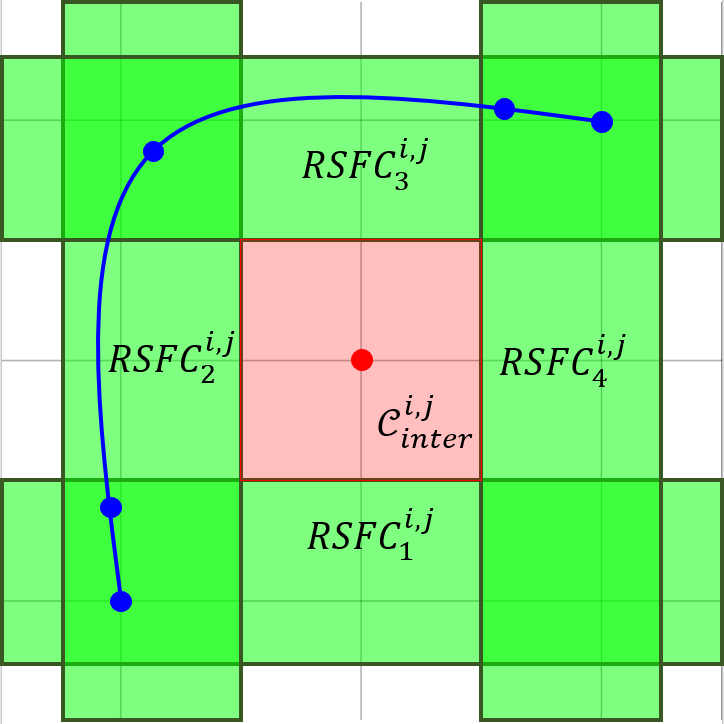}
        \caption{Example of Redundant RSFC transition.}
        \label{fig:rsfc redundant}
    \end{subfigure}
    \caption{Construction process of RSFC. For the convenience of explanation, it is depicted in 2D space. The initial trajectory and the relative initial trajectory of two multirotors are shown in Figs. \ref{fig:rsfc initial trajectory} and \ref{fig:rsfc relative trajectory}. The inter-collision model between two agents are depicted as a red box and RSFC is constructed along the waypoints of relative path avoiding the inter-collision model. Redundant RSFC transitions may occur as shown in Fig. \ref{fig:rsfc redundant}, so the greedy algorithm is used to minimize the number of RSFC transitions and the result is shown in Fig. \ref{fig:rsfc rsfc}.}
    \label{fig: rsfc construction}
\end{figure}

\subsection{Relative Safe Flight Corridor Construction}
As SFC models a obstacle-free space as a convex set, RSFC models a free space for evasive maneuver between two agents. We define RSFCs between the $i^{th}$ and $j^{th}$ agents, $RSFC^{i,j}_{1},...,RSFC^{i,j}_{M_{r}}$, as convex sets that do not invade the collision region between the $i^{th}$ and $j^{th}$ agents and are sequentially connected:
\begin{subequations}
for $m = 1,...,M_{r}$
\begin{align}
    RSFC^{i,j}_{m} \cap \mathcal{C}^{i,j}_{inter} = \emptyset
\label{eq: RSFC definition1}
\end{align}
and for $m = 1,...,M_{r}-1$
\begin{align}
    RSFC^{i,j}_{m} \cap RSFC^{i,j}_{m+1} \neq \emptyset
\label{eq: RSFC definition2}
\end{align}
\end{subequations}
where $\mathcal{C}^{i,j}_{inter}$ is a inter-collision model between $i^{th}$ and $j^{th}$ quadrotors which has a rectangular parallelepiped oriented with the body frame of $i^{th}$ quadrotor. Note that $\mathcal{C}^{i,j}_{inter}$ can vary for each pair of agents, which mean that it can handle different size of quadrotors. In our implementation, we assign length and width of $\mathcal{C}^{i,j}_{inter}$ as $2(r^i+r^j)$ and a height as $2c_{dw}(r^i+r^j)$ to consider downwash effect, where $c_{dw}$ is downwash coefficient.
The trajectory of $j^{th}$ quadrotor does not collide with $i^{th}$ quadrotor if for arbitrary $t \in [0,T]$, there exists  $m \in \{1,...,M_{s}\}$ such that $p^j(t)-p^i(t) \in RSFC^{i,j}_{m}$.


Construction of RSFC is described in Fig. \ref{fig: rsfc construction}. First, we convert the initial trajectory into the relative initial trajectory for each pair of agents (Fig. \ref{fig:rsfc relative trajectory}). The relative initial trajectory of $i^{th}$ and $j^{th}$ quadrotors, $p^{i,j}_{init}$, can be obtained by subtracting corresponding waypoints of two initial trajectories.
Next, for each waypoint in the relative path, we choose proper RSFC from RSFC candidates.
Using the quadrotors' differential flatness property, we design RSFC candidates to reduce the number of decision variable at the optimization step.
There are six RSFC candidates in direction $\pm x, \pm y, \pm z$, and each candidate $RSFC_\mu$ is defined as follows:
\begin{equation}
 RSFC_{\mu} = 
  \begin{cases} 
   \{p | p \cdot n_{\mu} > r^i+r^j\} & \text{if } \mu = \pm x, \pm y \\
   \{p | p \cdot n_{\mu} > c_{dw}(r^i+r^j)\} & \text{if } \mu = \pm z \\
  \end{cases}
\label{RSFC candidate}
\end{equation}
where $n_{\mu}$ is a unit vector in the direction $\mu \in \{\pm x, \pm y, \pm z\}$.
For each waypoint $p^{i,j}_{init}[k]$ in $p^{i,j}_{init}$, any $RSFC_{\mu}$ can be selected if $\mu$ satisfies the following condition:  
\begin{equation}
    p^{i,j}_{init}[k] \cdot n_{\mu} > 0
\label{eq: RSFC cand select}
\end{equation}
However, redundant RSFC transitions along the waypoints may increase the number of polynomial segments and the computation time. Figs. \ref{fig:rsfc rsfc} and \ref{fig:rsfc redundant} show the example. When we generate a smooth relative trajectory in RSFC, we need to plan two segment polynomials to represent the relative trajectory if there is one transition of RSFC (i.e. $RSFC^{i,j}_1 \rightarrow RSFC^{i,j}_2$) along the waypoints as shown in Fig. \ref{fig:rsfc rsfc}. However, if three transitions are involved (i.e. $RSFC^{i,j}_1 \rightarrow RSFC^{i,j}_2 \rightarrow RSFC^{i,j}_3 \rightarrow RSFC^{i,j}_4$) as shown in Fig. \ref{fig:rsfc redundant}, we have to plan two more polynomial segments compared to the previous one. Thus, we use the greedy algorithm (Alg. \ref{alg: RSFC construction}) to minimize the number of RSFC transitions.

\begin{algorithm}[h]
\caption{RSFC construction}
\label{alg: RSFC construction}
\begin{algorithmic}[1]
\State \textbf{Input:} $p^i_{init}$, $p^j_{init}$
\State $l_{max} \gets \max($ size($p^i_{init}$), size($p^j_{init}$))
\State $RSFC^{i,j} \gets \emptyset$ 
\ForAll{$\mu \in \{+x,+y,+z,-x,-y,-z\}$}  
    \State initialize $s_{\mu}[l_{max}]$ to 0 
\EndFor

\For{$n \gets 1$ to $l_{max}$}
    \ForAll{$\mu \in \{+x,+y,+z,-x,-y,-z\}$}
        \If {$(p^j_{init}[n] - p^i_{init}[n]) \cdot n_{\mu} > 0$}
            \If{n = 1}
                \State $s_{\mu}[n] \gets 1$
            \Else
                \State $s_{\mu}[n] \gets s_{\mu}[n-1]+1$
            \EndIf
        \EndIf
    \EndFor
\EndFor

\State $n \gets l_{max}$
\State $\mu_{M} \gets \arg\max_{\mu}(s_{\mu}[n])$
\State $RSFC^{i,j}.$push\_front$(RSFC_{\mu_{M}})$
\State $n \gets n - s_{\mu_{M}}[n]$
\While{$n > 0$}
    \State $\mu_{M} \gets \arg\max_{\mu \neq -\mu_{M}}(s_{\mu}[n])$
    \State $RSFC^{i,j}.$push\_front$(RSFC_{\mu_{M}})$
    \State $n \gets n - s_{\mu_{M}}[n]$
\EndWhile
\State \textbf{return} $RSFC^{i,j}$
\end{algorithmic}
\end{algorithm}

The algorithm receives $p^i_{init}$ and $p^j_{init}$ as input and returns $RSFC^{i,j}$. It initializes $RSFC^{i,j}$ (line 3) as an empty array and $s_{\mu}$ as an array of all zero with length $l_{max}$ (line 4-5). After initialization, the algorithm verifies RSFC candidates using (\ref{eq: RSFC cand select}) and saves the result in $s_{\mu}$ (line 6-12). 
At the end of the relative path, it finds an RSFC candidate that includes the maximum number of waypoints and appends the candidate to $RSFC^{i,j}$ (line 14-15). After that, it goes to the last waypoint among the included ones (line 16). Then again it finds the maximum including candidate until it reaches the start point of relative path (line 17-20). Note that the new candidate must not be located at the opposite side of the previous candidate because quadrotors cannot jump through an empty space between two opposite candidates (line 18).

\subsection{Time Segment Allocation}
\label{subsec: Time allocation}
To formulate an optimization problem, we determine the \textit{time segment} of the piecewise polynomial trajectory and allocate each SFC and RSFC to each polynomial segment. 
Let $p^i_{m}(t)$ be the $m^{th}$ segment of $p^i(t)$ which is defined at $t \in [t^i_{m-1}, t^i_{m}]$. The time segment of the $i^{th}$ quadrotor $t_s^i$ is defined as:
\begin{equation}
    t_s^i = [t^i_0,...,t^i_{M}]  
\end{equation}

In this paper, we set the trajectory of all agents to have the same time segment $t_s$ because it is necessary for our algorithm to utilize the convex hull property of Bernstein basis polynomial. However, it can result in too many decision variables, which can increase the computation time. Thus, the following method is used to decrease the number of decision variables. 

\vspace{1.5mm}
\begin{algorithm}[h]
\caption{Finding partial time segment}
\label{alg: partial time allocation}
\begin{algorithmic}[1]
\label{func: find partial ts}
\State \textbf{Input:} Initial or relative initial trajectory $p_{init}$, \\ \qquad \qquad Array of sequential convex sets $C$, \\ \qquad \qquad Time step $t_{step}$
\State $t_{sp} \gets \emptyset$
\State $m \gets 1$
\For{$n \gets 1$ to $l_{max}$}
    \If{$m \geq$ size$(C)$}
        \State break
    \EndIf
    \If{$p_{init}[n] \in (C[m] \cap C[m+1])$}
        \State $count \gets 1$
        \While{$p_{init}[n+count] \in (C[m] \cap C[m+1])$ \\ \qquad \qquad \textbf{and} $n+count \leq l_{max}$}
            \State $count \gets count + 1$
        \EndWhile
        \State $t_{sp}.$push\_back$(n+\floor[\big]{count/2}*t_{step})$
        \State $n \gets n+\floor[\big]{count/2}$
        \State $m \gets m+1$
    \ElsIf{$p_{init}[n] \in C[m+1]$} 
        \State $t_{sp}.$push\_back$((n+0.5)*t_{step})$
        \State $m \gets m+1$
    \EndIf
\EndFor
\State return $t_{sp}$
\end{algorithmic}
\end{algorithm}

To construct the time segment for all agents, we generate partial time segments $t_{sp}$ from SFC and RSFC. Alg. 2 shows the process of finding partial time segment. 
The algorithm receives SFC or RSFC and initial or relative initial trajectory as its input, and searches for the middle waypoint among the intersection of two sequential convex sets (line 11-13). After that, the algorithm records the index of this middle waypoint to assign as the location at which the SFC or RSFC transition occurs (line 14). In other words, $m^{th}$ SFC or RSFC is allocated before the time $(n+\floor[\big]{count/2})*t_{step}$ and $m+1^{th}$ SFC or RSFC is allocated after time $(n+\floor[\big]{count/2})*t_{step}$, where $n+\floor[\big]{count/2}$ is the index of middle waypoint among the intersection of $m^{th}$ and $m+1^{th}$ convex sets. In the SFC case, it is guaranteed that there exists a waypoint in two sequential SFCs because we connect the waypoint with SFC by the axis-search method. However, in the RSFC case, there may be no waypoint in an intersection between two sequential RSFCs (line 17).
In this case, we do not use the integer index because it can make an infeasible constraint when SFC and RSFC are changed simultaneously at the same time. Instead, we use a heuristic method that gives a time delay to RSFC transition to avoid the simultaneous change of SFC and RSFC (line 18). It may increase the number of the decision variables, but we can increase the success rate of finding a feasible trajectory. This algorithm always returns an array with a maximum size of $2l_{max}$, so it is guaranteed that the piecewise trajectory has a maximum of $2l_{max}-1$ segments.

After generating the partial time segments, we combine all of them into one and sort them. We delete duplicated elements and we generate the total time segment by appending the start time and total flight time at each end of the combined array. This method can reduce the size of the total time segment by overlapping the elements of partial time segment as much as possible. 

We allocate SFC and RSFC to time segment by comparing $t_{sp}$ with $t_s$. For example, assume that $t_s$ is determined as $[0, 1, 2, 3]$ and $t_{sp}$ of SFC is $[2]$. Then we can guess that $SFC^i_1$ is assigned for the first and second segments of piecewise polynomials and $SFC^i_2$ is assigned for the third segment. Let $SFC^i_{(m)}$ and $RSFC^i_{(m)}$ be convex sets that are allocated to the $m^{th}$ polynomial segment. In this example, SFC is allocated as $SFC^i_{(1)} = SFC^i_{(2)} = SFC^i_1$ and $SFC^i_{(3)}=SFC^i_2$.

\subsection{Trajectory Optimization}
\label{subsec: trajectory optimization}
In the optimization step, we plan the smooth polynomial trajectory using SFC, RSFC and time segment $t_s$, but it is difficult to handle SFC and RSFC with standard polynomial basis. Thus, we formulate the piecewise polynomial $p^i(t)$ of all agents as a piecewise Bernstein polynomial.

Bernstein basis polynomials of degree N are defined as:
\begin{equation}
    B_{k,N}(t) = {N\choose k}t^{k}(1-t)^{k}
\label{eq: bernstein basis}
\end{equation}
for $t\in[0, 1]$ and $k=0,1,...,N$, and Bernstein polynomial is the linear combination of Berstein basis polynomials.

The $m^{th}$ segment of $p^i(t)$ can be represented in Berstein polynomial as:
\begin{equation}
    p^i_m(t) = c^i_{m,0}B_{0,n}(\tau_m)+...+c^i_{m,N}B_{0,N}(\tau_m)
\end{equation}
where $\tau_m = \frac{t-t_{m-1}}{t_m-t_{m-1}}$ and $c^i_m = [c^i_{m,0},...,c^i_{m,N}]$ is the vector consisting of all control points of $p^i_m(t)$.

It is shown that a Bernstein polynomial has a convex hull property \cite{zettler1998robustness}, in other words, a Bernstein polynomial $p^i(t)$ is confined within the convex hull of its control points $c^i_m$. In \cite{tang2016safe,liu2017planning,gao2018online,honig2018trajectory}, it has been used to confine $p^i_m(t)$ within $SFC^i_{(m)}$ by limiting control point $c^i_m$ within $SFC^i_{(m)}$. 

Here, this convex hull property can be used to confine the relative polynomial trajectory. Assume that $p^i(t)$ and $p^j(t)$ have the same time segment, then the $m^{th}$ segment of $p^{i,j}_m$ can be written as:
\begin{equation}
\begin{split}
p^{i,j}_m(t) & = \sum\limits_{k=0}^{N}(c^j_{m,k}-c^i_{m,k})B_{0,n}(\tau_m)\\
             & = \sum\limits_{k=0}^{N}c^{i,j}_{m,k}B_{0,n}(\tau_m)
\end{split}
\end{equation}
where $c^{i,j}_{m,k} = c^j_{m,k}-c^i_{m,k}$ for $k=0,...,N$ is the control point of $p^{i,j}_m(t)$. We can observe that the relative Bernstein polynomial is also a Bernstein polynomial. Therefore, by the convex hull property, we can enforce quadrotors $i, j$ not to collide with each other by limiting all control points $c^{i,j}_{m,k}$ within $RSFC^i_{m}$. In this way, we can generate the safe trajectory by adjusting RSFC for each pair of agents. 

Our decision vector $c$ consists of all control points of $p^i_m(t)$ for $m = 1,...,M$ and $i=1,...,N_q$:
\begin{equation}
    c = [{c^1_1}^T,...,{c^1_M}^T,..., {c^{N_q}_1}^T,...,{c^{N_q}_M}^T]^T    
\end{equation}
where $M$ is the total number of polynomial segments that all agents share, and it is up to $2l_{max}$ as explained in section \ref{subsec: Time allocation}.

The cost function of polynomials is defined as follows:
\begin{equation}
    J = \sum\limits_{i=1}^{N_{q}}\sum\limits_{\mu\in\{x,y,z\} }\int_{0}^{T}(\frac{d^{n}p_{\mu}^{i}(t)}{dt^{n}})^2   
\label{eq: cost function}
\end{equation}
where $T$ is total flight time, and this can be represented into a quadratic form. In this paper, we set $n=3$, so that it minimizes the integral of the square jerk of total trajectory. It is a reasonable choice because we can minimize the input aggressiveness of quadrotor \cite{mueller2015computationally}. 

The waypoint constraints for start, goal positions and continuity constraints for smooth trajectory can be reformulated in linear equality constraints ($A_{eq}c=b_{eq}$). Therefore, our trajectory generation problem is reformulated as quadratic programming (QP) problem:
\begin{equation*}
\begin{aligned}
& \text{minimize}     & & c^{T}Qc \\
& \text{subject to}   & & A_{eq}c= b_{eq} \\
&                     & & c^i_{m,k}\in SFC^{i}_{(m)}, \forall i,k\\
&                     & & c^j_{m,k}-c^i_{m,k} \in RSFC^{i,j}_{(m)}, \forall i,j>i,k\\
\end{aligned}
\label{eq: optimization}
\end{equation*}
where $Q$ is the Hessian cost matrix derived by concatenating all Hessian cost function of individual agents with a block-diagonal matrix form. The detailed formulation of equality constraints can be found in \cite{gao2018online}. 
Note that we need only one QP to generate a smooth trajectory for all agents.

During optimization, we do not consider dynamic limits because they can be infeasible constraints for QP. Instead, we scale the time segment for all agents uniformly after optimization, similar to \cite{honig2018trajectory}.

\section{EXPERIMENTS}

\label{sec: experiments}
\begin{figure*}[t]
\centering
\includegraphics[width = 1\textwidth]{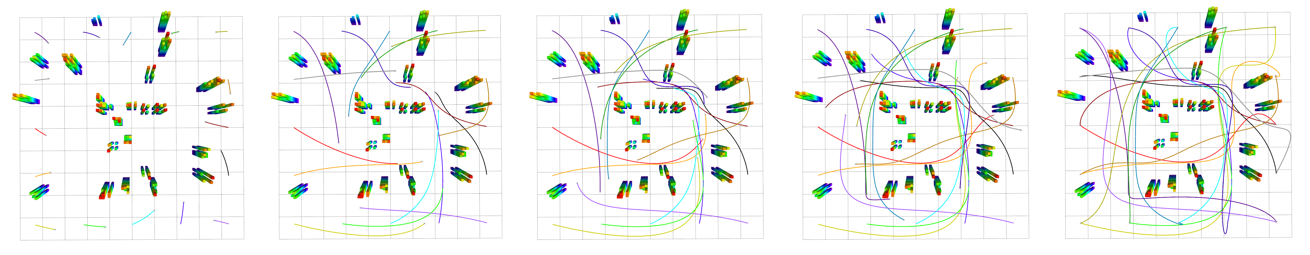}
\caption{Top-down view of 16 quadrotors in a 10 m $\times$ 10 m $\times$ 2.5 m random forest map. Goal points are opposite to start positions.}
\label{fig:simulation}
\end{figure*}

\subsection{Implementation Details}
We implement our proposed method in C++14. We use the Octomap library \cite{hornung2013octomap} to represent the 3D occupancy map and use the dynamicEDT3D library \cite{lau2013efficient} to compute distance between corridor and obstacle for SFC construction. For trajectory optimization, CPLEX QP solver \cite{cplex201612} is used to solve (\ref{eq: optimization}).

We model the collision models of quadrotors with radius $r = 0.15$m and downwash coefficient $c_{dw} = 2$ based on the specification of Crazyflie 2.0 in \cite{honig2018trajectory}. For initial trajectory planning, the grid size of the 3D grid map is determined to 0.5 m in x, y-axis directions and 1 m in z-axis direction. We set the degree of polynomials to $N = 5$ and give constraints to be continuous up to acceleration. Fig. \ref{fig:simulation} shows the planning result of 16 agents.

\subsection{Computation Time Evaluation}
\label{subsec: comp time}
We evaluate the computation time on a PC running Ubuntu 16.04. with Intel Core i7-7700 @ 3.60GHz CPU and 16G RAM. Our experiment is conducted in 10 m $\times$ 10 m $\times$ 2.5 m space. We randomly deploy 30 trees of size 0.3 m $\times$ 0.3 m $\times$ 1--2.5 m. Start positions of quadrotors are uniformly distributed in a boundary of the xy-plane in 1 m height, and we assigned the goal points at the opposite to their start position as shown in Fig. \ref{fig:simulation}.

\begin{table*}[t!]
\caption{Computation time by the number of quadrotors}
\label{table:scale}
\begin{center}
\begin{tabularx}{\linewidth}{c||*{4}{X}|X}
\toprule
Agents & \centering ECBS($c_w$=1.3) (s) & \centering SFC Construction (s) & \centering RSFC Construction (s) & \centering Traj. Optimization (s) & \centering Total Comp. Time (s) \tabularnewline
\hline 
4 &  \centering 0.034 & \centering 0.039 &  \centering 1.68E-5 & \centering 0.034 & \centering 0.11 \tabularnewline
8 & \centering 0.037 & \centering 0.053 & \centering 4.84E-5 & \centering 0.139 & \centering 0.23 \tabularnewline
16 & \centering 0.048 & \centering 0.081 & \centering 1.70E-4 & \centering 0.800 & \centering 0.93 \tabularnewline
32 & \centering 0.059 & \centering 0.137 & \centering 5.77E-4 & \centering 6.65 & \centering 6.86 \tabularnewline
64 & \centering 0.167 & \centering 0.256 & \centering 2.14E-3 & \centering 50.7 & \centering 51.2 \tabularnewline
\hline
\end{tabularx}
\end{center}
\end{table*}

We conduct the experiments by randomly changing the location of obstacles and measure the computation time of each step. Table \ref{table:scale} shows the average computation time of 30 experiments. The proposed method takes about a second for 16 quadrotors and a minute for 64 quadrotors. Although it uses the solver with $O(n^3)$ time complexity, the actual total computation time is short enough. 

\subsection{Success Rate Analysis}
In section \ref{subsec: Time allocation}, we give the time delay at the RSFC partial time segment to avoid infeasible constraints. To verify that,
we compared the two time allocation method, one is time allocation with RSFC time delay and the other is time allocation without the time delay by changing the line 18 of Alg. \ref{alg: partial time allocation} to $n*t_{step}$. We plan the trajectory of 16 quadrotors 50 times to measure the success rate. We use the same environment setting in section \ref{subsec: comp time} except quadrotor size. 

\begin{figure}[t]
\centering
\includegraphics[width = 0.4\textwidth]{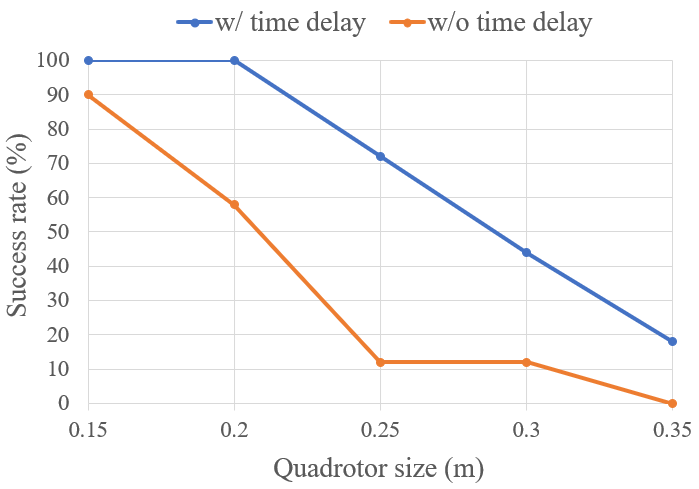}
\caption{Success rate of trajectory planning for 16 agents by the time allocation method.} 
\label{fig:success}
\end{figure}

Fig. \ref{fig:success} shows that the time allocation with RSFC time delay has a higher probability to find a feasible solution. It also shows a 100 percent success rate when quadrotor size is 0.15 m and 0.2 m. 
As expected, the success rate of both methods decrease as the quadrotor size increase.

\subsection{Flight Test}
\begin{figure}
    \centering
    \begin{subfigure}[t]{0.4\textwidth}
        \includegraphics[width=\textwidth]{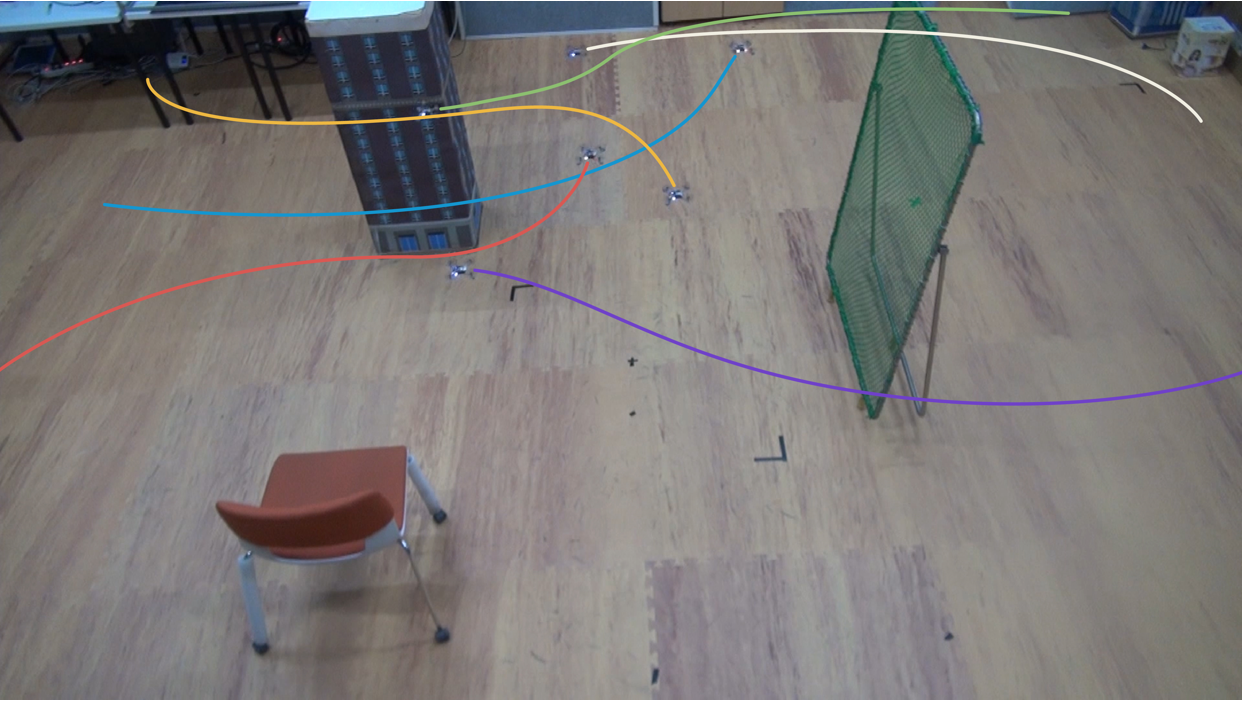}
        \label{fig:real snapshot}
    \end{subfigure}
    \begin{subfigure}[t]{0.4\textwidth}
        \includegraphics[width=\textwidth]{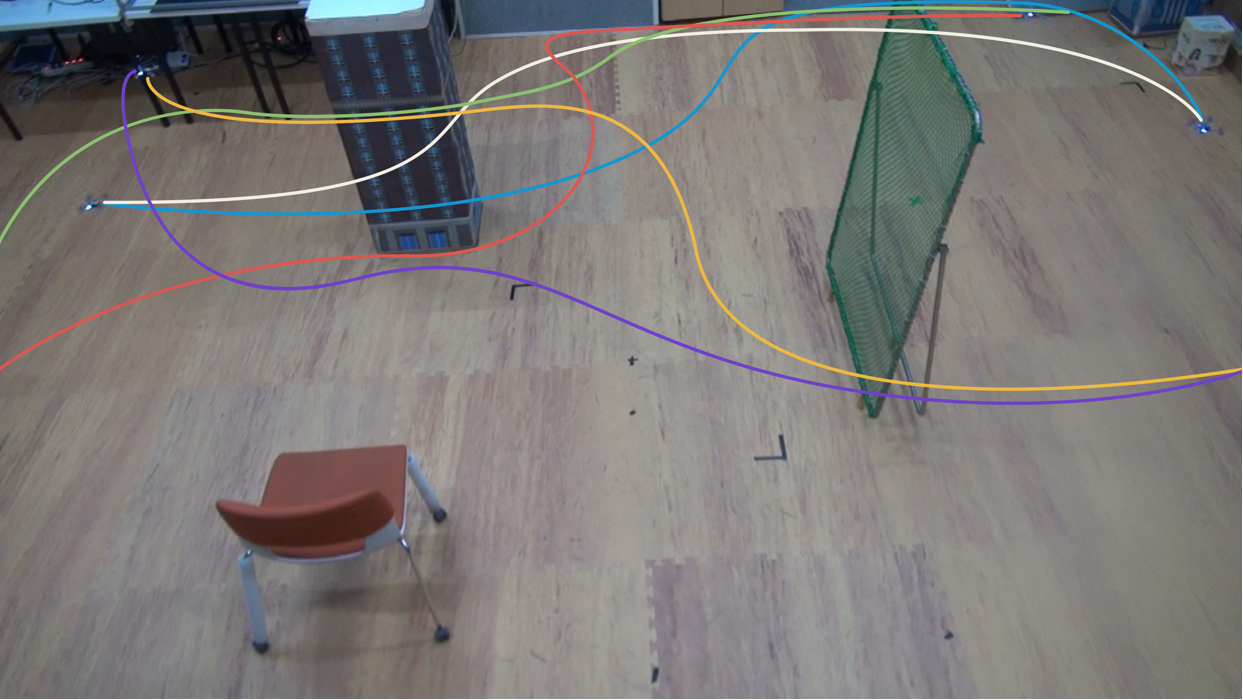}
        \caption{Snapshots of flight test.}
        \label{fig:real snapshot}
    \end{subfigure}
    ~ 
    \begin{subfigure}[t]{0.4\textwidth}
        \includegraphics[width=\textwidth]{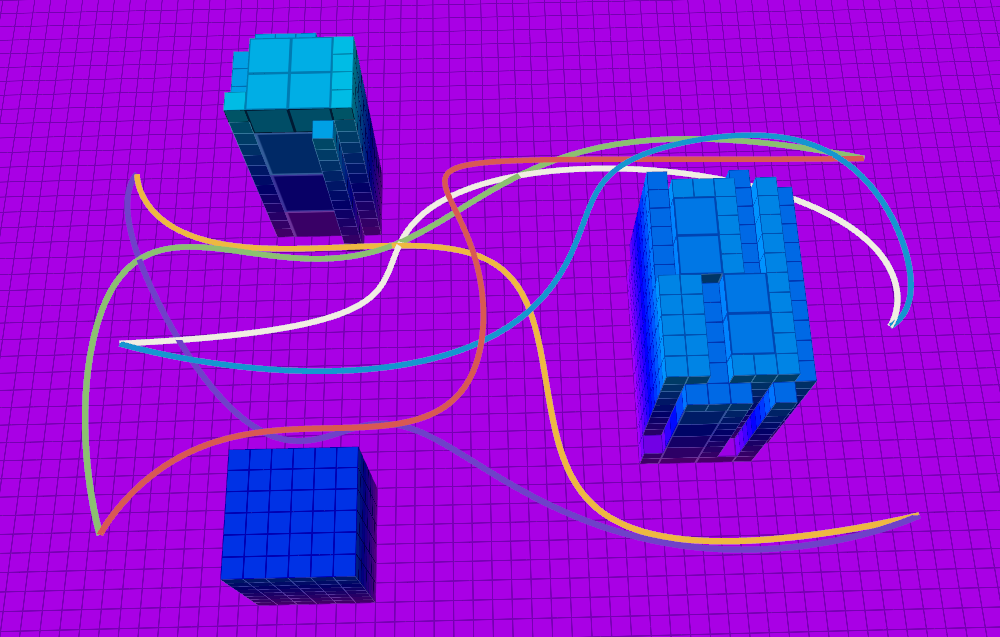}
        \caption{Trajectory plan for flight test.}
        \label{fig:real plot}
    \end{subfigure}
    \caption{Flight test with 6 quadrotors. Full flight is presented in the supplemental video.}
    \label{fig: real experiment}
\end{figure}

We demonstrate our algorithm with 6 Crazyflie 2.0 quadrotors in a 5 m x 7 m x 2.5 m space. Crazyswarm \cite{preiss2017crazyswarm} is used to follow the pre-computed trajectory, and Vicon motion capture system is used to estimate the position of each agent at 100 Hz.
It takes 0.138 seconds to plan the trajectory for all agents. Fig. \ref{fig: real experiment} shows the snapshot of flight test and the pre-computed trajectory. Full flight is presented in the supplemental video. 

\section{CONCLUSIONS}
\label{sec: conclusions}
In this paper, we propose a trajectory planning method using RSFC to deal with the inter-collision problem in a multi-MAV system. RSFC models the free space for inter-collision avoidance into a convex set, and we show that it can be converted into linear constraints by utilizing the convex hull property of Bernstein polynomial. To generate trajectory for multiple quadrotors, we adopt the ECBS algorithm to obtain the initial trajectory in a 3D grid map. Then we construct SFC and RSFC based on the initial trajectory, and allocate them to each segment considering infeasible simultaneous transition. Finally, an optimization solver generates a smooth trajectory that is collision-free and deadlock-free. The proposed method can generate a safe trajectory for 64 agents in a minute, and flight test is executed to validate our solution.  

In future work, we plan to reduce the computational effort of our work for online trajectory generation and we plan to develop more precise time allocation method that guarantees a feasible solution.

\newpage
\printbibliography

\addtolength{\textheight}{-12cm}   




\section*{ACKNOWLEDGMENT}
This material is based upon work supported by the Ministry of Trade, Industry \& Energy(MOTIE, Korea) under Industrial Technology Innovation Program. No.10067206, ‘Development of Disaster Response Robot System for Lifesaving and Supporting Fire Fighters at Complex Disaster Environment’

This work was supported by the Robotics Core Technology Development Project (10080301) funded by the Ministry of Trade, Industry and Energy (MoTIE, Korea)


\end{document}